# How to estimate the association between change in a risk factor and a health outcome?

M Katsoulis[1,2], AG Lai[1,2], DK Kipourou[3], R Sofat[1,2], M Gomes[4], A Banerjee[1,5,6], S Denaxas[1,2,7] RT Lumbers[1,2], K Tsilidis[8,9], H Hemingway[1,2], KD Ordaz[10]

[1]Institute of Health Informatics, University College London, London, UK

[2]Health Data Research UK, University College London, London, UK.

[3]Inequalities in Cancer Outcomes Network, Department of Non-communicable Disease Epidemiology, London School of Hygiene and Tropical Medicine, UK

[4]Department of Applied Health Research, University College London, London, UK

[5]University College London Hospitals NHS Trust, London, UK

[6]Barts Health NHS Trust, The Royal London Hospital, London, UK

[7]Alan Turing Institute, London, UK

[8]Department of Epidemiology and Biostatistics, School of Public Health, Imperial College London, London, UK

[9]Department of Hygiene and Epidemiology, University of Ioannina School of Medicine, Ioannina, Greece

[10]Department of Medical Statistics, London School of Hygiene and Tropical Medicine, UK

**Standfirst**

Estimating the effect of a change in a particular risk factor and a chronic disease requires information on the risk factor from two time points; the enrolment and the first follow-up. When using observational data to study the effect of such an exposure (change in risk factor) extra complications arise, namely (i) "when is time zero?" and (ii) "which information on confounders should we account for in this type of analysis? From enrolment or the 1$^{st}$ follow-up? Or from both?". The combination of these questions has proven to be very challenging. Researchers have applied different methodologies with mixed success, because the different choices made when answering these questions induce systematic bias. Here we review these methodologies and highlight the sources of bias in each type of analysis. We discuss the advantages and the limitations of each method ending by making our recommendations on the analysis plan.

**BOX**

- It is challenging to estimate the effect of a change in a risk factor between enrolment and the 1st follow-up on a health outcome using observational data. It is not straightforward to define "baseline" and then adjust for "baseline confounders"

- In this paper, we show that among the different methods used in the literature to tackle this problem, the less problematic (and, thus, recommended) method implicitly defines a baseline period, instead of a single date for baseline. Within this period, time zero should be at the date of the measurement of this risk factor at enrolment and change in this risk factor will be defined by the change in the measurements between baseline and the 1st follow-up.

- Researchers should adjust for confounders (or exclude individuals based on levels of confounders) measured both at enrolment and at 1st follow-up.

- Two important problems remain unsolved:

  a) Usually the change in the risk factor can be caused by different interventions, which researchers cannot specify from their dataset

  b) Researchers have to assume that individuals in their study do not die or do not develop the outcome of interest till the 1st follow-up in which the risk factor is measured for the 2nd time (immortal time bias)

# INTRODUCTION

In studies that use observational data from cohort or Electronic Health Records (EHR), it is challenging to estimate the effect of an exposure which is defined as a change in a risk factor between enrolment and the 1$^{st}$ follow-up on the disease of interest. Firstly, it is not straightforward to define "baseline", should it be the date of enrolment or the date of 1$^{st}$ follow-up? Secondly, the appropriate use of information on the confounders is non-trivial. Researchers have used different methods to tackle this problem; however, the results arising from these methodologies are expected to differ, because different choices lead to different sources of systematic biases. Using a simple hypothetical trial as an example and assuming that we had access to part of these data from an observational database, we discuss in detail these methodologies, and highlight the possible sources of bias for each one of them. We further discuss the strengths and the limitations of each and conclude with a recommendation on the analysis plan on how to assess the association between a change in a risk factor and a health outcome using observational data.

# HOW TO ESTIMATE THE ASSOCIATION BETWEEN CHANGE IN A RISK FACTOR AND A HEALTH OUTCOME

The majority of research articles focuses on the relationship between a risk factor and a health outcome[1]. The conclusion from this type of analysis can be important for prediction purposes, e.g. high levels of BMI are associated with increased risk for fatal and non-fatal cardiovascular disease (CVD); yet, this does not translate to a specific course of action. Ideally, a randomised control trial would be conducted, in which individuals with overweight or obesity would be assigned into diet or physical activity regimens. A given follow up period would be defined during which the CVD events would be collected. Since such randomised trials are rare, expensive and difficult to conduct, there is an increasing interest in using observational studies [2-14]. The situation is further complicated as data on physical activity or diet are typically not recorded. However, questions such as "What is the effect on cardiovascular disease of bodyweight reduction?" can be answered using EHR or data from consented cohorts [2-14]. Undeniably, to assess the effect of weight/BMI change using observational data requires information on individuals' bodyweight for two time points; at study enrolment and the 1st follow-up. This means that the exposure of interest is only defined at the first follow-up point. This introduces ambiguity to the definition of "baseline", where eligibility criteria determining who is selected into the study should be met, as well as to the choices of confounders that should be controlled for, as "baseline" confounders are intrinsically related to related to the definition of "baseline". In other words, should we only include confounders corresponding to enrolment, the 1st follow-up, or both?

This article discusses these issues, with a special emphasis on the definition of "baseline". To illustrate the problem and build intuition, we present a simplified example, motivated by obesity research, though conclusions will apply to other fields.
We first demonstrate results from two hypothetical trials, before moving onto assessing the same associations, using with the same individuals identified from a healthcare database. We apply different methodologies that have been reported in the literature, compare the findings and assessing the sources of bias in each case. Finally, we will discuss the advantages and the limitations of assessing the results of these studies and we will make our suggestions on the analysis plan, which comes from the literature of target trial emulation using observational data [12-14]

**Hypothetical trials**

Consider two hypothetical trials with perfect adherence to the protocol and no loss to follow-up. Both trials enrol individuals aged 45-60 years, free of chronic diseases at baseline. The 1st trial enrols individuals with BMI from 25 to 29.9kg/m$^2$, commonly referred to as overweight individuals, while the 2$^{nd}$ enrols individuals with BMI 30 to 34.9kg/m$^2$, commonly referred to as individuals with obesity I. In each trial, participants are randomly assigned into two treatment arms: the 1$^{st}$ arm consists of high intensity physical activity and low caloric intake for 2 years. These combinatorial interventions is expected to result in ~5kg/m$^2$ BMI loss; thus, this is considered the BMI loss arm of the trial. The 2$^{nd}$ arm includes high intensity physical activity and standard caloric intake for 2 years. This is expected to result in no BMI changes; thus, this is considered the BMI maintenance arm. The protocol allows individuals to abandon their intervention if they develop a chronic condition (e.g. cancer, renal failure etc) during the intervention period (2 years). Individuals are then followed up for another 18 years (20 years in total). The endpoint is defined as the occurrence of a fatal and non-fatal cardiovascular disease during the follow-up. For more details, see table 1. Since our focus is on discussing the sources of systematic biases, rather than statistical considerations relating to estimation and precision, we imagine that we have a very large sample, and in particular, each individual in table 1 as representing millions of individuals with the same data, so that the 95% confidence intervals around the point estimates are very narrow.

In this simple example, it is clear from Table 1 that (i) there is no difference between the two interventions on CVD after 20 years in the overweight, and (ii) the 1$^{st}$ intervention (low caloric intake and high levels of physical activity, which caused BMI loss) resulted in a lower risk of CVD compared to the 2$^{nd}$ intervention. More specifically, the cumulative risk of CVD is 3/20=15% in both arms in the overweight, while in the obese, the CVD risk is 4/20=20% in the weight loss arm and 5/20=25% in the weight maintenance arm. From this (hypothetical) example, 10% (2/20) of the participants (in both arms, both in the overweight and the obese) developed a clinically condition (e.g. cancer, diabetes etc) which allowed them to deviate from their intervention till the 1$^{st}$ follow-up. In other words, these participants developed a chronic disease independently of their intervention. Among those who developed a chronic disease, all of them lost weight, independently of the arm they were initially randomised (both in the overweight and the obese) and then 50% of them developed CVD till the end of follow-up.

**Observational data**

Now, let us assume that we would like to conduct the same analysis using a cohort from a healthcare database, which has exactly the same people with the above-mentioned hypothetical trials (Table 1), pooled together. However, information on both diet and physical activity are limited in such data. Most of these databases have information on physical activity, but relatively few on diet. So, for simplicity, assume that we know all the information contained in Table 1, apart from dietary intake on our observational cohort. We could try to partially recover this information, by taking into consideration BMI change in the 1$^{st}$ follow-up (after two years). However, there are two extra problems, compared to the usual analysis of observational data; (i) weight or BMI change is not a well-defined intervention[12,13,15] and (ii) the definition of "baseline" is non-trivial, which subsequently complicates the analysis plan for the selection of confounders.

To further simplify the exposition, consider a situation where the only factor that affects BMI change between baseline and the 1$^{st}$ follow-up, apart from physical activity and diet, is the occurrence of a chronic disease [either the outcome (CVD) or other chronic conditions like cancer, diabetes etc].

Next, we analyse these data using 4 different methods proposed in the literature and highlight the pitfalls and the differences in the analysis. We present the protocol of the hypothetical trials and the observational study in Table 2.

1. *a. Baseline definition: Time zero is defined at enrolment, i.e. the time where all eligibility criteria should be met. Individuals are allocated to a BMI change group based on their BMI at enrolment and their observed weight/ BMI trajectories (from enrolment till the 1$^{st}$ follow-up)*

   *b. Baseline confounders: Confounders measured at enrolment*

This approach implicitly assumes[2-3] that since time zero is defined just before the beginning of the exposure (i.e. BMI change), all confounders can be accounted for either by adjusting for them or excluding participants based on the existence of specific chronic diseases at enrolment according to the eligibility criteria (see figure 1). This would be the correct strategy for the well-defined interventions, when we are interested in emulating trials with time-fixed interventions using observational data (e.g. when we focus on the effect of bariatric surgery[4]). Yet, when the exposure is less well-defined and it is expressed as a change in a risk factor at 2 time points, e.g. weight/ BMI change, this approach may lead to severe bias as we will present below.

Weight change observed between enrolment and the 1$^{st}$ follow-up visit may be due to intermediate events (such as developing cancer) that occurred during this period. In our illustrative example, this approach estimates that, in the overweight, the risk of CVD is 4/22~18.2% in the BMI maintenance arm and 2/18~11.1% in the BMI loss arm. Moreover, in the obese, the risk of CVD is 5/22=22.7% in the BMI maintenance arm and 4/18~22.2%% in the BMI loss arm. Therefore, instead of estimating a null effect in the overweight (result from the RCT), we obtain a risk difference (BMI loss vs maintenance) equal to 7.1%, while in the obesity cohort, instead of a beneficial effect of BMI loss vs maintenance (estimated at 5% from the RCT when comparing the BMI loss vs maintenance arm of specific interventions), we estimate a slightly adverse effect (risk difference ~0.5%). These differences in the findings occurred because this strategy fails to take into account that people's observed BMI loss might be due to a confounder (e.g. chronic disease) between the chosen time zero and the 1st follow-up.

2. *a. Baseline definition: Time zero is defined at the time of the 1st follow-up where all eligibility criteria are also met. Individuals are allocated to a BMI change group based on their BMI at the 1st follow-up and their observed weight/ BMI trajectories (from enrolment till the 1st follow-up)*
   *b. Baseline confounders: Confounders measured at the 1st follow-up*

There are many studies that have adopted this approach to estimate the relationship between weight/BMI change and a health outcome[5-8]. However, the estimand (i.e. the parameter of interest we end up estimating) here is awkward. Using this approach, this parameter of interest is the effect of weight change amongst individuals at a certain BMI group, measured after the (hypothetical) intervention (weight change), i.e. at the 1st follow-up visit. In our example, we cannot estimate the effect of BMI loss vs maintenance in those with obesity, because there are no individuals in that group after the BMI loss intervention. In the overweight cohort (figure 3), this strategy results in a risk difference (weight loss vs maintenance) equal to 5.6%, instead of estimating a null effect (result from the RCT).

As a result, this method can only be used for prediction purposes, i.e. in risk scores to account for history of weight/BMI trajectories. Any causal interpretation based on these definitions is highly problematic.

3. *a. Baseline definition: Time zero is defined at the 1st follow-up, in which all eligibility criteria related to chronic diseases are met. Individuals are allocated to a BMI change group based on their BMI at enrolment and their observed weight/ BMI trajectories (from enrolment till the 1st follow-up)*

   *b. Baseline confounders: Confounders measured at the 1st follow-up*

Given that time zero is defined 2 years after the enrolment, in these settings, the follow-up time will be 18 years (20-2=18 years). This approach[9-11] has many advantages compared to the previous two. First, this strategy correctly excludes individuals who developed chronic diseases from enrolment till the end of the 1st follow-up, therefore removing confounding (in this simplified example); these individuals had not necessarily been "allocated" (which in the observational dataset means observed) to the BMI loss intervention at baseline (e.g. this was the case for individuals 21, 22, 61 and 62). Moreover, we have successfully allocated individuals to a BMI change group based on their initial BMI (at enrolment) and their observed BMI trajectory (from enrolment till the 1st follow-up). In the overweight group (see Figure 1 and 2), this strategy estimates that there is no risk difference between weight loss and weight maintenance (as in the trial). Moreover, it estimates (correctly) a beneficial effect of weight loss (vs weight maintenance) in the obesity group.

However, the concern is that the time-zero (1st follow-up) is not aligned with the beginning of the exposure (enrolment). This method has led many researchers to adjust for the participant's characteristics (apart from BMI at enrolment) corresponding to the time of the 1st follow-up, which is considered time zero[9-11]. Using this approach, usually these studies ignore other variables occurring at or before enrolment, but are indeed associated with weight/ BMI at enrolment (and therefore with exposure) and the outcome (see Figure S2 in the Appendix). Furthermore, there may be some bias introduced by excluding those with chronic diseases between enrolment and 1st follow-up, if these diseases are developed as a consequence of weight change.

We see that the magnitude of the absolute risks and risk differences has small difference compared to the true risks (see Figure 2 and 3). Specifically, when comparing the risk differences from the observational study and the trial, there is the estimates are the same in the overweight and very similar in the obese (differed by 0.6%). It is of note though that this comparison is not well defined, because the estimand (i.e. the quantity we aim to estimate) is different in the analysis of the trial and the observational database. In the trial, we measure the effect of physical activity and diet on CVD, while from the observational data, the quantity of interest is the effect of BMI change on CVD. However, in our simplified scenario, in which BMI change can be caused either by physical activity, diet or a chronic disease (i.e. we assume that no individual was on orlistat or other drugs, no individual on chopped off her arms etc), all individuals had the same levels of physical activity and we can account for (i.e. exclude individual with) chronic disease in the analysis of observational data, then any BMI change is due to the different diet in the two arms.

In general, we expect this bias to be relatively small, if a) the chronic diseases occurring from enrolment till the 1st follow-up are not caused by the weight change b) the period between enrolment and the 1st follow-up is relatively small, compared to total follow-up time (e.g. <15%) and c) the incidence of chronic diseases up to 1st follow-up is relatively small (e.g. <15%). Moreover, if the aforementioned assumptions hold, the direction of risk differences will be correct if the chronic diseases which occurred between the enrolment and the 1st follow-up are not related to the weight change intervention.

4. *a. Baseline definition: Time zero is defined at enrolment, thus coinciding with the beginning of the intervention. Individuals are allocated to a BMI change group based on their weight or BMI at enrolment and their observed weight/ BMI trajectories from enrolment till the 1$^{st}$ follow-up (which define the intervention). Instead of a single time point for baseline, it's used a baseline period. Since the exposure (hypothetical intervention) is only known at the first follow up period, we need to adjust for confounding, both at enrolment, and at the 1$^{st}$ follow-up. Eligibility criteria for chronic diseases should be met during the period between enrolment and 1$^{st}$ follow-up.*

   *b. Baseline confounders: Confounders at enrolment and at the 1$^{st}$ follow-up*

This methodology has been used on more recent papers that use the causal inference framework to mimic hypothetical interventions[12-14] and is the one we recommend. These studies usually focus on sustained interventions over time (not only till the 1$^{st}$ follow-up), however, in this paper, we just adopt their definition of baseline.

The main difference between this method and method 3 is that the follow-up time in method 4 will be 20 years, as it begins at enrolment. Since eligibility criteria for chronic conditions must be met from enrolment until we observed the exposure (at 1$^{st}$ follow up), we must exclude those that develop the outcome, or any other chronic disease. This means that our estimand only applies to people experiencing "healthy" weight change during the first 2 years, that do not develop the outcome. In our example we have only one confounder (chronic disease at 1$^{st}$ follow-up) based on the values of which we exclude individuals. Usually, we additionally need to adjust for other confounders in our outcome regression models[13], to emulate randomisation in the baseline period.

The absolute risk and the risk differences resulting from this methodology will be the same as those from methodology 3.

Our recommended approach has two advantages, compared to methodology 3:

a. It is consistent with the timing of the interventions. Time zero is defined at the beginning of the interventions (i.e. at enrolment).
b. Most importantly, to emulate randomisation at baseline using method 3 in cohort databases or EHR, the eligibility criteria for chronic conditions needs to be correctly specified at the end of the 1$^{st}$ follow-up; however, there is a potential risk that the adjustment for "baseline" confounders is not adequate. For example, researchers who follow method 3 usually adjust for the participant's characteristics (apart from BMI at enrolment) only at the time of the 1$^{st}$ follow-up[9-11], which is considered time zero. On the other hand, method 4, requires that the eligibility criteria for chronic conditions are met from time-zero (enrolment) till the end of the 1$^{st}$ follow-up and therefore, we recommend it. Additionally, researchers should adjust for i) all the confounders measured at enrolment (e.g. age, sex, weight or BMI at enrolment, prevalence of hypertension, use of diuretics prior to enrolment etc) and ii) all the additional confounders measured at the 1$^{st}$ follow-up, i.e. variables that are potentially associated with BMI/weight change (the observed intervention), e.g. smoking cessation, use of diuretics between enrolment and 1$^{st}$ follow-up etc. These adjustments are essential because we only observe the exposure corresponding to each individual after time-zero; It is obvious from figure S1 in the Appendix when is the time of recording of the exposure, the confounders and the outcome, while it is obvious from figure S2 the need of adjusting for confounders at time zero to block any back-door pathway between the exposure and the outcome.

On the other hand, a potential concern with method 4 is that it assumes that there is no risk of developing the outcome for the first 2 years, this introduces immortal time bias[17], In other words, individuals are assumed to be alive and not develop the outcome till the first follow-up. The bias resulting from this method is not expected to be substantial, if (a) the chronic diseases occurred from the enrolment till the 1st follow-up are not caused by the weight change interventions , (b) the period between enrolment and the 1st follow-up is relatively small, compared to total follow-up time and (c) the incidence of chronic diseases between enrolment and the 1st follow-up is relatively small.

**DISCUSSION**

In this article, we highlighted the challenges arising when estimating the effect of an exposure, defined as a risk factor's change, on a health outcome. Researchers have used different methodologies for defining baseline, deciding which confounder measurements to adjust for as well as the defining the eligibility criteria for inclusion. We assessed these biases and the problematic interpretations from the analysis using four different methodologies and discussed why method 4, which has been recently proposed within a causal inference framework to emulate hypothetical interventions[12-14], is the most adequate. This recommended approach involves defining time zero at enrolment and identifying individuals to a BMI change group based on their eligibility criteria for chronic diseases being satisfied from enrolment till the 1st follow-up. Thus, in this method, it is used a baseline period, from enrolment till the 1st follow-up, and we proposed to adjust for all potential confounders (or exclude individuals, based on the eligibility criteria), measured at time zero, as well as during this baseline period.

**Strengths and Limitations**

Using the recommended method 4, it enables us to minimise the risk of residual confounding by not adjusting for confounders measured both before enrolment and during the period where the exposure is not yet observed (between enrolment and 1st follow-up). Researchers following method 3 usually adjusted for the participants' characteristics (apart from BMI at enrolment) only at the time of the 1st follow-up[9-11], which is considered "time zero". As we show in Figures 2 and 3, using our proposed methodology, the direction of the association between a risk factor's change and a health outcome is correct, while there is only small bias in the magnitude of this relationship. Moreover, it is consistent with the timing of the interventions as time zero is defined at the beginning of the interventions.

This study has some limitations. Weight/ BMI change is not a well-defined intervention[12,13,15]. In other words, we do not know how people lost, maintained or gained weight (consistency assumption), as we assumed data on specific weight-changing strategies (such as diet and exercise regimes) are not available in EHR databases, thus limiting a more specific definition of a hypothetical weight-change intervention. Unfortunately, none of the 4 methodologies described can address this. Therefore, our estimates should be interpreted as the combined effects of weight loss strategies (exercise, restricted-caloric intake, drugs etc) that result in weight changes. This problem appears in other setting, when, for example, the exposure is change in systolic blood pressure[17]. Moreover, it should be noted that the results from the observational data from all these methods are expected to differ from the results of the (hypothetical) trial, because the estimand is different. In the trials, the causal contrast is the comparison of well-defined interventions, defined by physical activity and caloric intake (and assuming that everybody adhered). Using the observational data, the causal contrast is the comparison of BMI change trajectories, which are a consequence of hypothetical interventions. Additionally, we assumed there is no information that allows us to distinguish intentional (healthy) from unintentional weight changes, other than developing a chronic disease between time zero and 1st follow up, which is accounted for in methods 3 and 4.

Moreover, it is clear from our (hypothetical) dataset that none of the strategies above is unbiased. Using the 1st or 2nd approach, neither the magnitude nor the direction of the risk differences (weight loss vs maintenance) was correctly estimated; needless to say, the conclusions for these analyses would be completely wrong. Our recommended (4th) methodology tackles the problem of the dependence of a risk factor's change from a chronic disease that occurred between enrolment and the first follow-up. Yet, we face the problem of immortal time bias [16] as a result of all individuals needing to survive and remain healthy till the first follow up period, to be included in the study. Unlike other examples, in which the problem of immortal time bias can be tackled through cloning (or through randomly assigning the individual to one of the strategies)[18-19], this technique cannot be applied in paradigms regarding BMI/weight change, because individuals should be compatible with only one hypothetical intervention at enrolment (under perfect adherence), so that they end up either losing or maintaining weight.

**Conclusion**

It is very important to define a baseline period between enrolment and the 1st follow-up, when the exposure of interest is a change in a risk factor. To account for "baseline confounders", one needs to adjust for (or exclude for, based on the eligibility criteria) confounders both at enrolment and the 1st follow-up.

**FUNDING**

MK is funded by the British Heart Foundation (grant: FS/18/5/33319).

KDO is supported by UK Wellcome Trust Institutional Strategic Support Fund- LSHTM Fellowship 204928/Z/16/Z.

SD is supported by an Alan Turing Fellowship.

AGL is supported by funding from the Wellcome Trust (204841/Z/16/Z), National Institute for Health Research (NIHR) University College London Hospitals Biomedical Research Centre (BRC714/HI/RW/101440), NIHR Great Ormond Street Hospital Biomedical Research Centre (19RX02) and the Health Data Research UK Better Care Catalyst Award.

HH is a National Institute for Health Research (NIHR) Senior Investigator. His work is supported by: 1. Health Data Research UK (grant No. LOND1), which is funded by the UK Medical Research Council, Engineering and Physical Sciences Research Council, Economic and Social Research Council, Department of Health and Social Care (England), Chief Scientist Office of the Scottish Government Health and Social Care Directorates, Health and Social Care Research and Development Division (Welsh Government), Public Health Agency (Northern Ireland), British Heart Foundation and Wellcome Trust. 2. The BigData@Heart Consortium, funded by the Innovative Medicines Initiative-2 Joint Undertaking under grant agreement No. 116074. This Joint Undertaking receives support from the European Union's Horizon 2020 research and innovation programme and EFPIA; it is chaired, by DE Grobbee and SD Anker, partnering with 20 academic and industry partners and ESC. 3. The National Institute for Health Research University College London Hospitals Biomedical Research Centre.

Table 1: Hypothetical randomised trial in which 80 people* are randomly assigned to one of two intervention: a) low caloric intake and high physical activity and b) standard caloric intake and high physical activity. Same data were analysed from an observational database, in which information on levels of caloric intake was missing (in red)

| ID | number of individuals | BMI group | level of caloric intake | Physical activity | BMI change (in 2 years follow-up) | BMI group after 2 years | chronic disease (in 2 years follow-up) affecting BMI change | CVD (end of follow-up, i.e. at 20 years) |
|---|---|---|---|---|---|---|---|---|
| 1 | 1 | overweight | Low | high | loss | normal weight | yes | yes |
| 2 | 1 | overweight | Low | high | loss | normal weight | yes | no |
| 3-4 | 2 | overweight | Low | high | loss | normal weight | no | yes |
| 5-20 | 16 | overweight | Low | high | loss | normal weight | no | no |
| 21 | 1 | overweight | Standard | high | loss | normal weight | yes | yes |
| 22 | 1 | overweight | Standard | high | loss | normal weight | yes | no |
| 23-24 | 2 | overweight | Standard | high | remain | overweight | no | yes |
| 25-40 | 16 | overweight | Standard | high | remain | overweight | no | no |
| 41 | 1 | obese | Low | high | loss | overweight | yes | yes |
| 42 | 1 | obese | Low | high | loss | overweight | yes | no |
| 43-45 | 3 | obese | Low | high | loss | overweight | no | yes |
| 46-60 | 15 | obese | Low | high | loss | overweight | no | no |
| 61 | 1 | obese | Standard | high | loss | overweight | yes | yes |
| 62 | 1 | obese | Standard | high | loss | overweight | yes | no |
| 63-66 | 4 | obese | Standard | high | remain | obese | no | yes |
| 67-80 | 14 | obese | Standard | high | remain | obese | no | no |

*As mentioned in the text, each of these individuals may represent millions (i.e. precision in the confidence interval is beyond the scope of this paper)

Table 2: Protocol of the hypothetical trial and the hypothetical observational study

|  | Trial that we would like to do but is not feasible | Observational study ||||
|---|---|---|---|---|---|
|  |  | Methodology 1 (no adjustment for confounders at 1st follow-up) | Methodology 2 (Initial BMI measured after BMI change) | Methodology 3 (Follow-up starts after BMI change) | Methodology 4 (proposed) |
| **Research question*** | What is the effect of being allocated to a specific intervention defined by physical activity and caloric intake (i.e. intention-to-treat) | what is the effect of weight change either from a chronic disease or from different hypothetical interventions | what is the effect of healthy weight change (as a result of different hypothetical interventions) among individuals at a certain BMI group, measured after the interventions | what is the effect of healthy weight change (as a result of different hypothetical interventions) ||
| **Eligibility criteria** | Trials would enrol otherwise healthy individuals at baseline, aged 45-60yo. Individuals with prevalent CVD or other chronic diseases at date of enrolment are exclude. The trials will be conducted in a) overweight and b) obese individuals. | Same criteria applied at enrolment | Same criteria applied at the 1st follow-up | Same criteria applied at the 1st follow-up | Same criteria applied both at enrolment and at the 1st follow-up |
| **Treatment strategies** | a) High physical activity and moderate caloric intake (expecting result: BMI loss of 5kg/m$^2$ in 2 years) b) High physical activity and high caloric intake (expecting result: BMI maintenance in 2 years) These interventions would last for 2 years. | a) BMI loss b) BMI maintenance These interventions would last for 2 years. ||||
| **Time zero (beginning of follow-up)** | At enrolment | At enrolment | At 1st follow-up || At enrolment |
| **Allocation to interventions**** | Physical activity and caloric intake are used at enrolment | BMI is used at enrolment and BMI change at the 1st follow-up | Both BMI and BMI change are used at the 1st follow-up | BMI is used at enrolment and BMI change at the 1st follow-up ||
| **Assignment procedures** | Random assignment to treatment strategy | Non-randomly assigned to a BMI change intervention ||||
| **Follow-up** | 20 years | 20 years | 18 years || 20 years |
| **Outcome** | Fatal and non-fatal CVD | Same | Same | Same | Same |
| **Statistical analysis** | None | Apply exclusion criteria at enrolment. Adjust for confounders measured at enrolment in the outcome regression models (in our example from table 1, we have no such confounders). | Apply exclusion criteria at 1st follow-up. Adjust for confounders measured at the 1st follow up in the outcome regression models. || Apply exclusion criteria both at baseline and 1st follow-up. Adjust for confounders measured both at enrolment and the 1st follow-up in the outcome regression models. |

*The research question reflects the estimand in these studies, i.e. the quantity we aim to measure
**The allocation to a specific intervention is based on baseline definition

Figure 1: Summary of different methods used to estimate the effect of a change in a risk factor (in our example; BMI change) and a health outcome (in our example; CVD) from an observational study. Definition of baseline, adjustment for baseline confounders, criteria for individuals' allocation to specific groups as well as for exclusion from the study

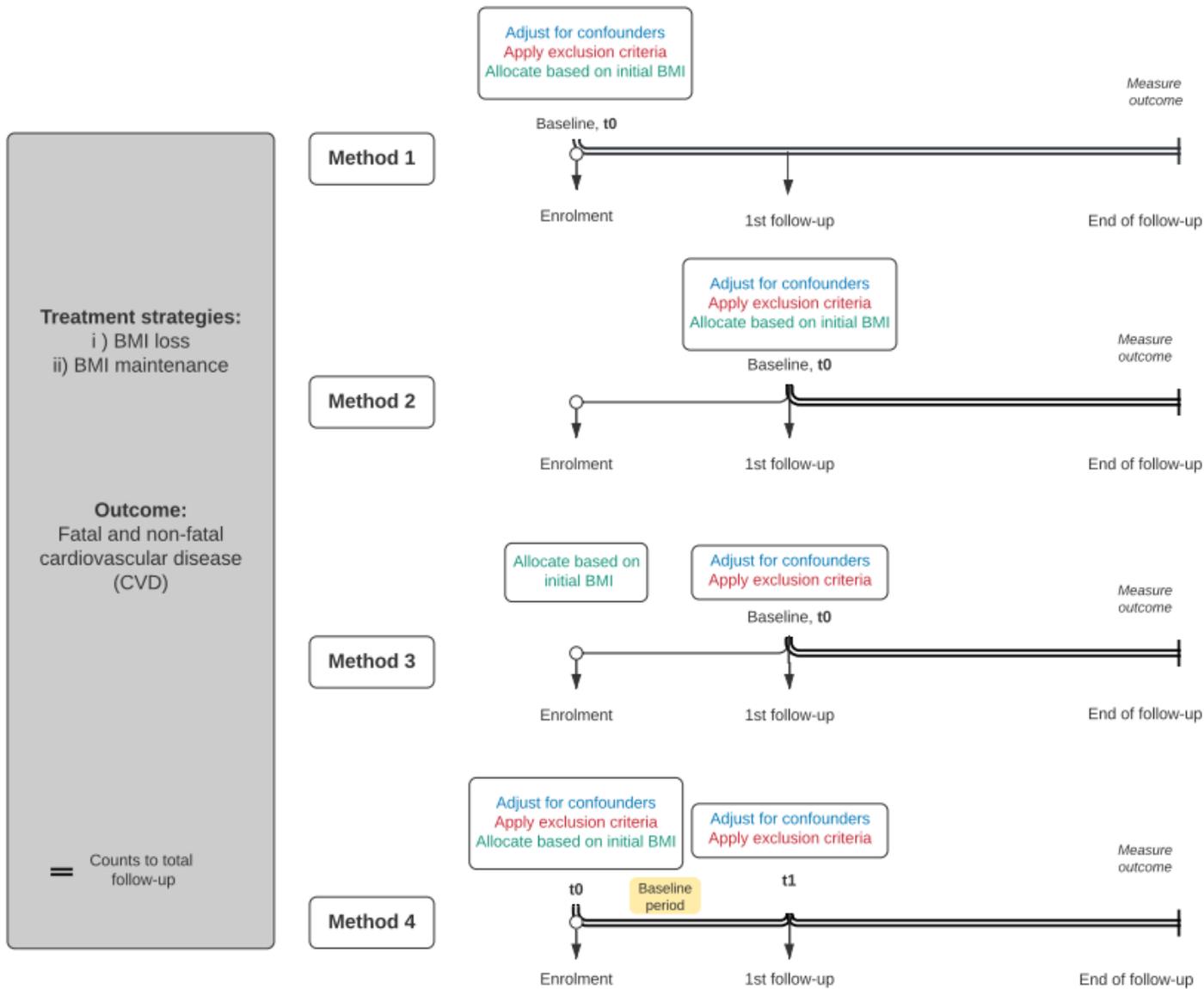

Figure 2: Flowchart showing how risk for developing the outcome (fatal and non-fatal CVD) is calculated using methods‡ 1, 3 and 4.

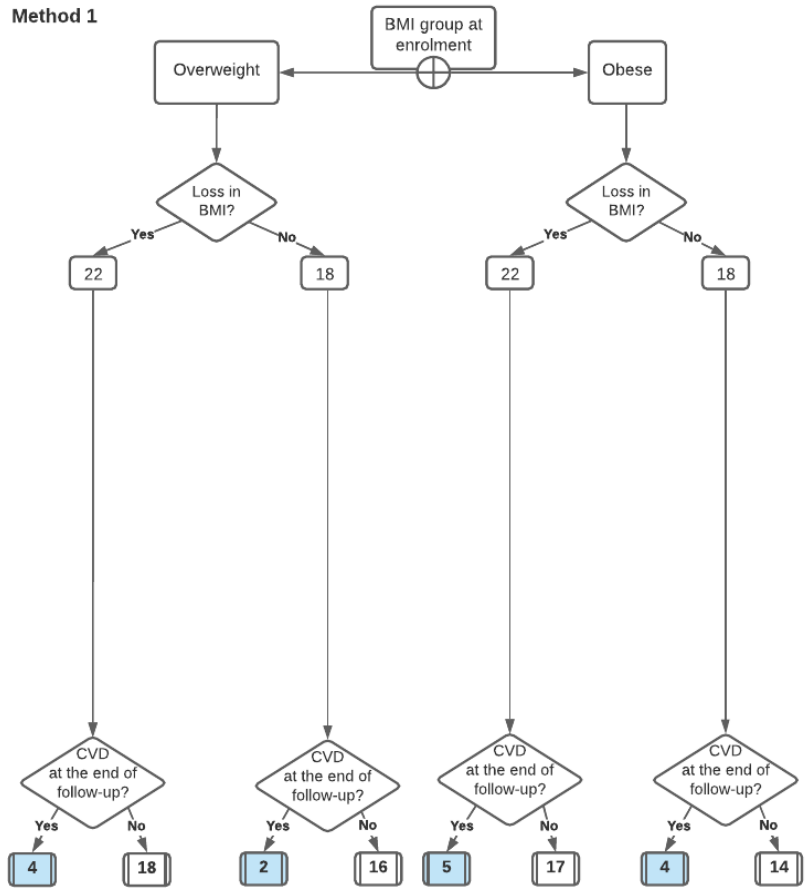
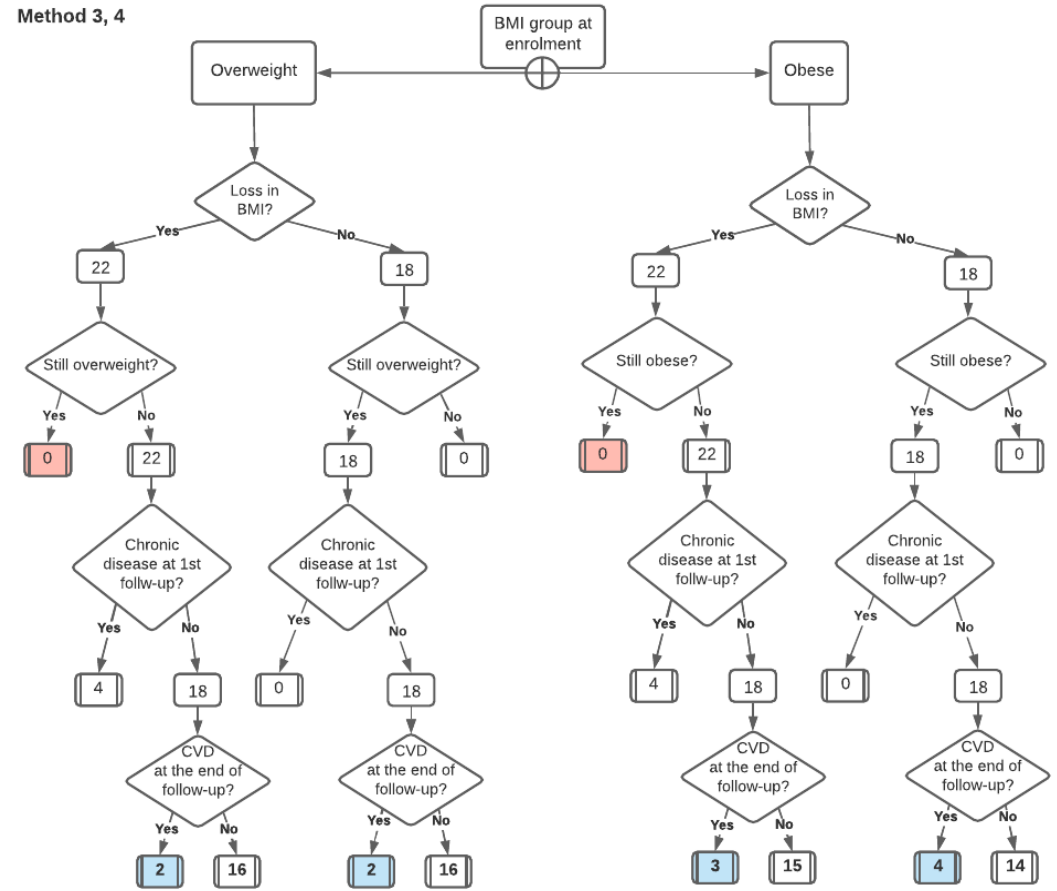

‡method 2 is not presented because allocation to a BMI group at baseline is done after BMI loss/ maintenance

Figure3: Estimation of risk difference of BMI loss vs BMI gain on fatal and non-fatal CVD from our hypothetical example in overweight and obese individuals (see Table 1), when using different methods to estimate the effect of BMI change, and comparison† with the corresponding risk difference of the interventions in the trial (red line)

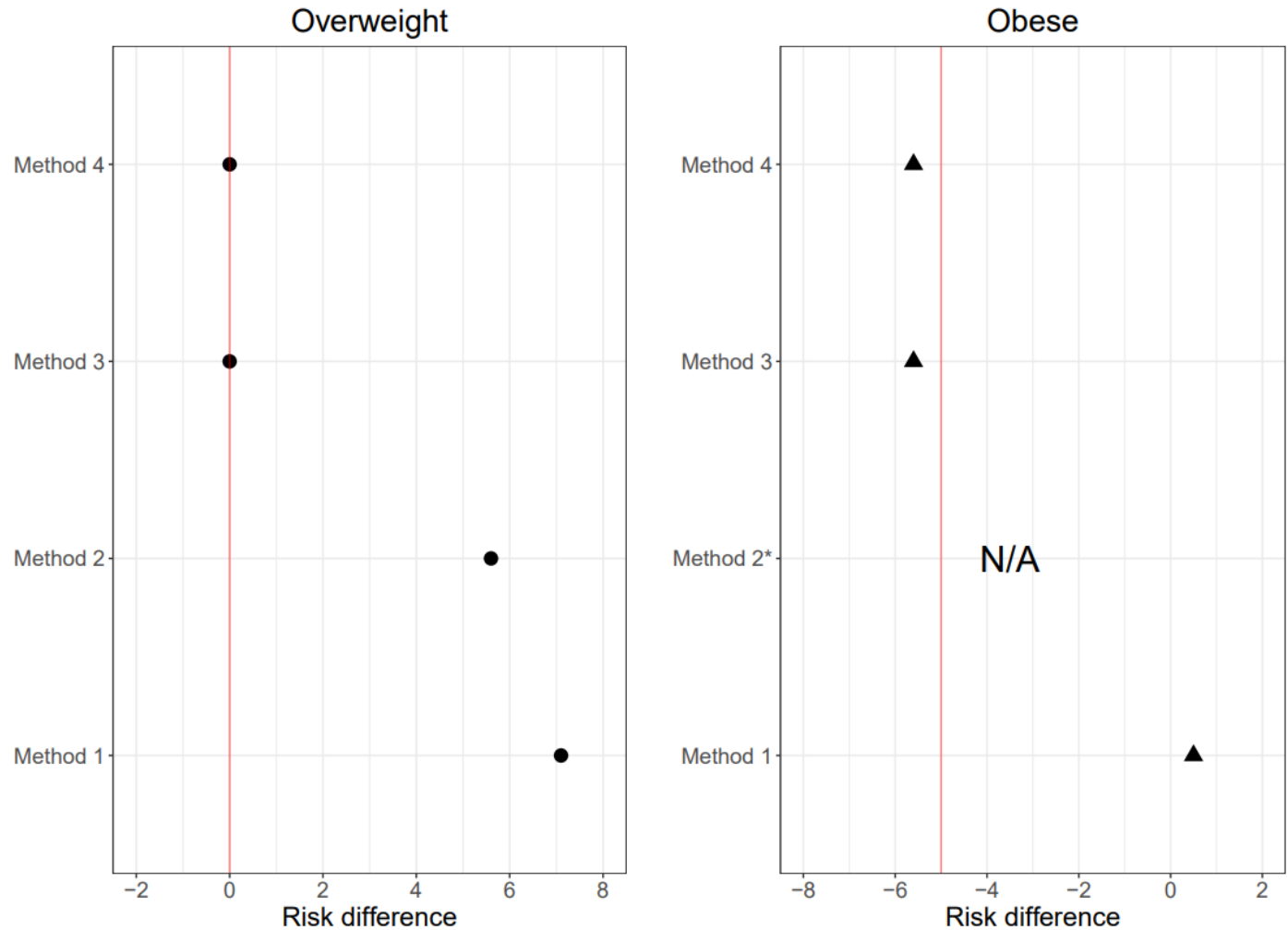

†This comparison is not well defined, because the estimand (i.e. the quantity we aim to estimate) is different in the analysis of the trial and the observational database, even if we talk for the same individuals. In the trial, we measure the effect of physical activity and diet on CVD, while from the observational data, the quantity of interest is the effect of BMI change on CVD. However, in our oversimplistic scenario, in which BMI change can be caused either by physical activity, diet or a chronic disease (i.e. no individual on orlistat or other drugs, no individual on chopped off her arms etc), and all individuals had the same levels of physical activity and we can account for (i.e. exclude individual with) chronic disease in the analysis of observational data, then the BMI loss and BMI maintenance arm of the trial are closely related to the BMI change from the observational data

*In method 2, we cannot estimate the bias, because we cannot estimate the risk difference in the obese; individuals with obesity are identified after the intervention. So, there are no obese individuals in the BMI loss group

# APPENDIX

Figure S1: Data from BMI/weight measurements, confounder[*] and outcome (CVD) needed to estimate the association of weight change in the first 2 years with CVD (from Table 1)

|  | T=0 | T=1 | T=2 | T=3 | T=4 | T=5 | T=6 | T=7 | T=8 | T=9 | T=10 |
|---|---|---|---|---|---|---|---|---|---|---|---|
| **Years from time zero (enrolment)** | 0 | 2 | 4 | 6 | 8 | 10 | 12 | 14 | 16 | 18 | 20 |
| **BMI measurements** | $B_0$ | $B_1$ | | | | | | | | | |
| **Confounder**[*] | $C_0$ | $C_1$ | | | | | | | | | |
| **Allocation to a BMI group** | | ⬇ | | | | | | | | | |
| **Outcome (CVD) measured**[**] | $O_0$ | $O_1$ | $O_2$ | $O_3$ | $O_4$ | $O_5$ | $O_6$ | $O_7$ | $O_8$ | $O_9$ | $O_{10}$ |

[*]In table 1, there is only one confounder (chronic disease). Physical activity is another confounder, but all individuals have the same (high) levels of physical activity. In method 4, we exclude individuals with a chronic disease (in alliance with the eligibility criteria)

[*]Using method 4, we exclude individuals if they had prevalent CVD at enrolment ($O_0=1$) or developed the outcome in the time until the 1st follow-up ($O_1=1$)

Figure S2: Directed acyclic graph (DAG) for the effect of BMI change on CVD in observational studies.

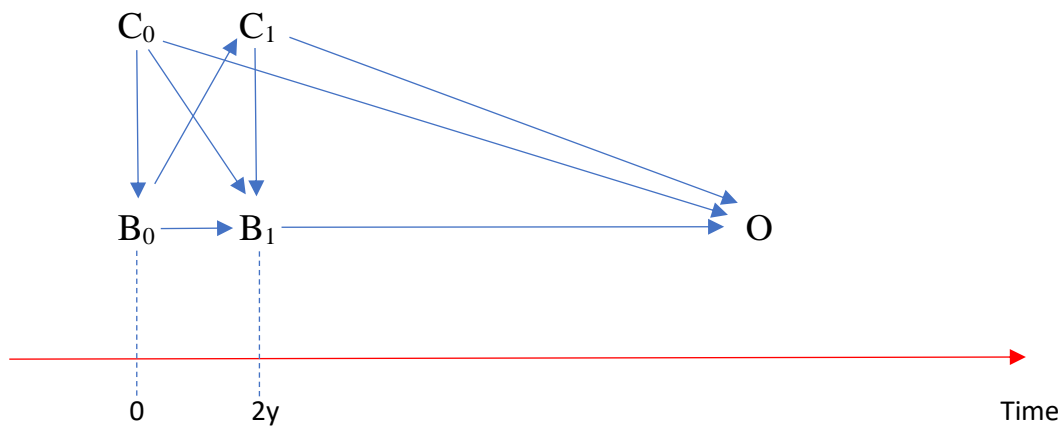

The confounders at time 0 (enrolment) $C_0$ and time 1 (1st follow up) $C_1$ (variables which simultaneously associated with BMI at either time and CVD) need to be controlled for. BMI change is only observed at time 1, i.e. when we can measure $B_1-B_0$. In other words, for standard levels of initial BMI $B_0$, BMI loss or maintenance can be observed only from $B_1$. From this DAG, it is clear that should control for both $C_0$ and $C_1$. If we do not control for $C_0$, we leave open the backdoor pathway
$B_1$<--$C_0$--> O open. In the same fashion, if we do not control for $C_1$, we leave open the backdoor pathway $B_1$<--$C_1$--> O open.